\documentstyle[epsf]{l-aa}  
\def\ph1{\phantom{1}}
\def\phu{\phantom{.0}}
\begin{document}
\thesaurus{03(02.18.5; 13.18.1)}
\title{Synchrotron spectra and ages\\
of compact steep spectrum radio sources}
%
%
%
  \offprints{M. Murgia}
  \date{Received 7 January 1999 / Accepted 12 March 1999}           
\author{M. Murgia\inst{1,2} \and C. Fanti\inst{1,3} \and R. Fanti\inst{1,3} \and
L. Gregorini\inst{1,3} \and U. Klein\inst{4} \and K.-H. Mack\inst{1,4} \and M. 
Vigotti\inst{1}}  
\institute{Istituto di Radioastronomia del CNR,
               Via Gobetti 101, I-40129 Bologna, Italy \and
           Dipartimento di Astronomia,
               Universit\`a di Bologna,
               Via Zamboni 33, I-40126 Bologna, Italy \and
           Dipartimento di Fisica,
               Universit\`a di Bologna,
               Via Irnerio 46, I-40126 Bologna, Italy \and
           Radioastronomisches Institut der Universit\"at Bonn,
               Auf dem H\"ugel 71,
               D--53121 Bonn, Germany 
}
%
\maketitle
\markboth{Ages of CSS sources}{Murgia et al.}
\begin{abstract}
 The high-frequency integrated spectra of Compact Steep Spectrum (CSS) sources  
show breaks with a moderate spectral steepening well fitted by continuous
injection synchrotron spectra. In lobe-dominated CSS sources the radiative ages
deduced by the synchrotron theory are in the range of up to $10^5$ years, if
equipartition magnetic
fields are assumed. These radiative ages are well correlated with the
source size indicating that {\it the CSS sources are young}.
 In order to maintain the {\it frustration scenario}, in which the sources' 
lifetimes are $\approx 10^7$ years, their equipartition magnetic field would
be systematically decreased by a factor $\ga 20$.
To complete the sample used in this work, we conducted observations at
230~GHz with the IRAM~30-m telescope of those sources which did not have such 
high-frequency observations up to now.
\end{abstract}

\section{Introduction}
Flux density limited radio source samples are known to contain a large
portion of compact sources with high frequency spectral index $\alpha$
$\ge$ 0.5 (S$_{\nu} \sim \nu^{-\alpha}$) and  angular sizes below 2$''$. The
corresponding projected linear sizes are typically $\le$ 15~kpc\footnote{We
assume H$_0$ = 100 km~s$^{-1}$~Mpc$^{-1}$ and q$_0=0.5$}.
 Fanti et al. (1990) have shown that the majority of these CSS sources
 cannot be larger sources foreshortened by projection 
effects, which means that their radio emission originates on 
sub-galactic scales. 
                                                                     
It is obvious that one should investigate the connection between CSS
and larger radio sources. Two scenarios have been proposed that
would naturally explain the observed small sizes. First, they could 
reflect an early stage in the evolution of radio sources. This is the
{\it youth scenario} (Phillips $\&$ Mutel 1982; Carvahlo 1985).
The second possibility is that the unusual conditions in the 
interstellar medium of their host galaxies, such as a higher density
and/or the presence of turbulence, inhibit the radio source from growing to 
larger sizes. This is the {\it frustration scenario} (van Breugel et al. 1984).
                                                                        
Most CSS sources exhibit double-lobed structures such as seen in classical
radio galaxies. This feature is common to both quasars and galaxies.
Quite a few of these are symmetric, which gave rise to the terms CSO
(compact symmetric objects, with sizes $\le$ 0.5~kpc) and MSO
(medium-size symmetric objects, with sizes $>$ 0.5~kpc). These sources
are considered as scaled-down versions of larger-sized double radio
sources (Fanti et al. 1995). A minority of CSS sources is made up by sources
with complex or highly asymmetric structures. These are mostly quasars,
with most of their luminosity provided by the jet. The distortions may
be caused by jet bending, which is further amplified by strong
projection. \par \noindent
A plausible organization scheme for these different morphological classes is to
identify CSO and MSO with progenitors of large doubles (`baby radio galaxies'),
while the asymmetric CSS sources could represent frustrated radio sources.

\section{Sample selection}
The sample we choose for this study consists of: \par \noindent
1) the sample selected from the 3CR and the Peacock \& Wall (1981; 
PW) catalogues (Fanti et al. 1995), and \par \noindent
2) nine sources extracted from the B3\,VLA sample (Vi\-gotti et al. 1989). 

1) The 3CR/PW sample (38 sources) was established on the basis of the well-known
criteria: linear size $<$ 15 kpc, spectral index $\alpha >$ 0.5 and a limit in
radio power P $\geq 10^{26.8}$ W/Hz at 178 MHz for the 3CR, or P $\geq 10^{26}$
W/Hz at 2.7 GHz for the PW.
High resolution structure information is available for these sources from the 
literature. In the framework of this project we observed at 230~GHz
the 17 sources for which the flux densities at this frequency were
not available from literature.

2) The second complete sample of CSS sources was selected from the ``strong'' 
section of the B3\,VLA catalogue ($S_{408 {\rm MHz}}\geq$ 0.8 Jy) according
to the following definitions: $\alpha_{408{\rm MHz}}^{1.4{\rm GHz}} > 0.5$
up to 10~GHz, and a size of $\la$ 15~kpc. Sources with $\alpha_{408{\rm
MHz}}^{1.4{\rm GHz}}< 0.5$ have also been included if their spectrum is steeper
than $\alpha=0.5$ at $\nu>1.4$~GHz. These are Giga-Hertz-peaked
spectrum (GPS) candidate sources. This sample contains 83 objects.
The sample has been recently observed with the VLA at 5 and
8.4 GHz, with resolutions of $\sim 0\farcs2$. Additional observations with
VLBI arrays are planned.
Only about 60\% of these sources have at the moment either a photometric or a 
spectroscopic redshift, therefore we limited the list to the objects which 
could be observed at 230~GHz to the latter. 
Due to scheduling constraints we only observed 9 sources of the B3\,VLA sample.

\begin{figure*}
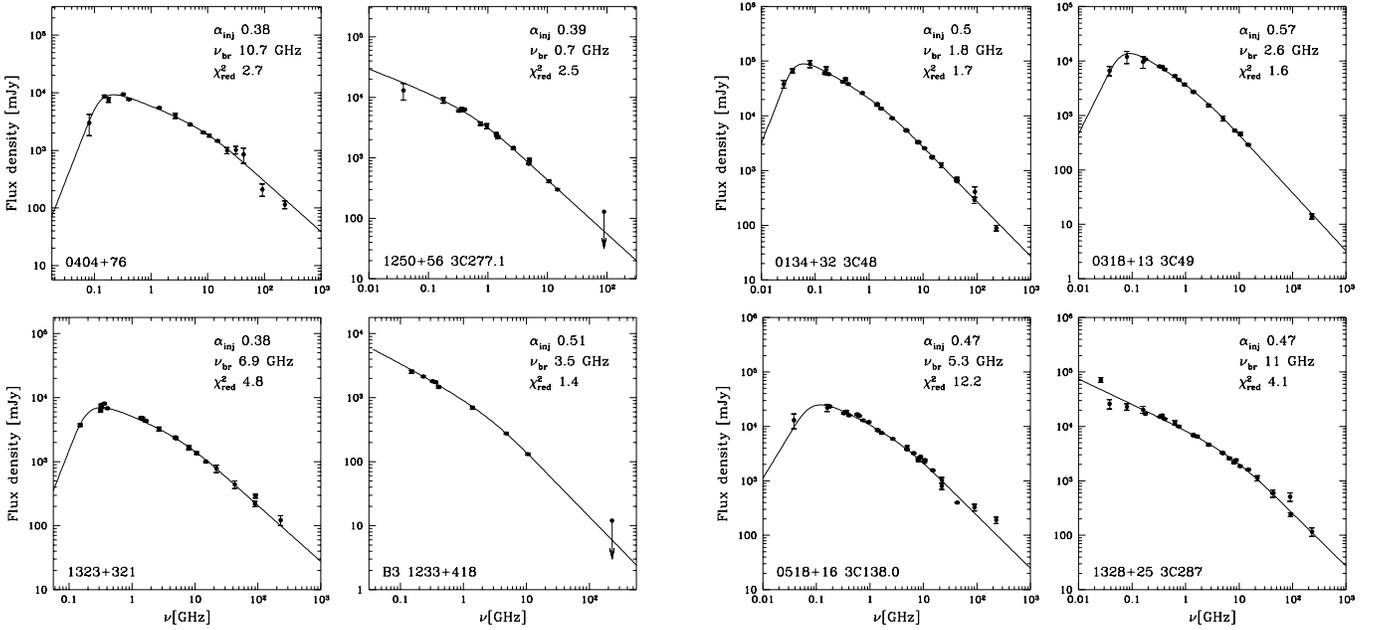

\begin{minipage}[h]{8cm}
\epsfxsize=8cm
\epsfbox  [45 190 540 720] {8461.f1a}
\end{minipage}
\hfill
\begin{minipage}[h]{8cm}
\epsfxsize=8cm
\epsfbox  [80 190 575 720] {8461.f1b}
\end{minipage}
\caption[]{Selected source spectra (left: type a, right: type b; see Sect. 4.3). The
$\chi^2_{\rm red}$-values refer to the pure synchrotron loss models. The
self-absorbed part of the spectrum has been modelled in a subsequent fit
procedure.}
\end{figure*}

\section{Observations and data reduction}
The observations were done with the IRAM~30-m telescope near Granada, Spain
between 10th Feb. and 15th Feb., 1998. The receiver we used was the 37-channel
bolometer described by Kreysa et al. (1998), with  feeds arranged in a
hexagon around the central feed, and beam separations of 20$\arcsec$. The
sensitivity of each channel was 70\,mJy/s. The beam size was
$10\farcs5$ as derived from pointing scans. All our sources were point-like to 
this beam. Opacity measurements were made via
skydips, from which we derived zenith opacities of 0.1--0.8. Despite the high
atmospheric attenuation of the signal, the observations (generally done at
high source elevations) could be performed satisfactorily, owing to the
relatively quiet atmosphere present during our observing run (resulting in 
a relatively low sky noise). The calibration factor to convert the observed 
counts into flux densities was determined by observing the source K3--50A,
which has a total flux density of 10\,Jy and a peak flux density of 7.6\,Jy 
for the beam size of our measurements. The measures were finally corrected for 
atmospheric attenuation and the elevation-dependent gain of the telescope. 
The pointing was frequently checked by cross-scanning the sources 0923+392,
1144+402, 1308+326, and 1418+546. 
The pointing accuracy was found to be $\sim\,3\arcsec$. The same sources were 
used to adjust the focus at regular intervals.

The sources were observed in an ON-OFF mode in which the wobbling secondary
mirror of the telescope moved at a frequency of 2\,Hz between the source 
position and positions located at $\pm~60\arcsec$ in azimuth. Each source
was observed between 3 and 5 times, with 10 ON-OFF pairs each. The differences
of the flux densities between the main and the reference beams were averaged by
weighting inversely proportional to the square of the r.m.s. noise of each 
measurement. 
The resulting integration time spent on each source was thus typically 50 
minutes, and the final r.m.s. noise of each measurement was between 2 and 
5\,mJy. In addition, the calibration may be uncertain by 15\% owing 
to day-to-day variations in the system gain.

The data were recorded with all receivers, with the central one pointing
at the source. As the angular extents of the target sources are 
in the range of a few arcseconds, the outer channels
contain only sky emission. Hence, they were used to subtract a mean sky level
from the central channel. Using this procedure, we assumed that the scale size
of the fluctuations of the sky emission is large compared to the total
coverage of the bolometer array ($60\arcsec$), which corresponds to 1.5\,m at
a distance of 5\,km. Together with the wobbling rate, this ensures
efficient rejection of sky noise, since the time scale of atmospheric
fluctuations is much larger (see e.g. Altenhoff et al. 1987).

In Tab.~1 we list the measured flux densities and 1-$\sigma$ errors, inclusive
of noise and calibration uncertainty, at 230~GHz.

\begin{table}
\caption{Flux densities at 230 GHz}
\centering
\begin{tabular}{|lccr|}\hline
 Name & \multicolumn{2}{c}{Pointing centre (B1950)} & 
\multicolumn{1}{c|}{S$_{230}$}\\
      & \multicolumn{1}{c}{$\alpha$}& \multicolumn{1}{c}{$\delta$}   & 
\multicolumn{1}{c|}{[mJy]}     \\
\hline
 3C-PW          &             &             &                   \\
0404+76         & 04 04 00.13 & 76 48 52.50 &  115 $\pm$     18 \\
0428+20         & 04 28 06.86 & 20 31 09.13 &   62 $\pm$     11 \\ 
0740+38 3C186   & 07 40 56.77 & 38 00 30.97 &    5 $\pm$  \ph14 \\
0758+14 3C190   & 07 58 45.05 & 14 23 04.40 &    8 $\pm$  \ph13 \\ 
1005+07 3C237   & 10 05 22.04 & 07 44 58.56 &   14 $\pm$  \ph15 \\
1019+22 3C241   & 10 19 09.38 & 22 14 39.63 &    2 $\pm$  \ph13 \\
1153+31         & 11 53 44.08 & 31 44 16.00 &    9 $\pm$  \ph13 \\
1203+64 3C268.3 & 12 03 54.08 & 64 30 18.40 &    1 $\pm$  \ph12 \\
1225+36         & 12 25 30.77 & 36 51 47.00 &    0 $\pm$  \ph13 \\
1358+62         & 13 58 58.36 & 62 25 06.71 &   29 $\pm$  \ph16 \\
1416+06 3C298   & 14 16 38.80 & 06 42 21.30 &    8 $\pm$  \ph14 \\
1447+77 3C305.1 & 14 47 49.35 & 77 08 46.65 &    2 $\pm$  \ph13 \\
1517+20 3C318   & 15 17 50.63 & 20 26 52.95 &    5 $\pm$  \ph13 \\
1634+62 3C343   & 16 34 01.06 & 62 51 41.80 &   12 $\pm$  \ph14 \\
1637+62 3C343.1 & 16 37 55.29 & 62 40 34.24 &    7 $\pm$  \ph14 \\
1819+39         & 18 19 42.33 & 39 41 15.00 &    5 $\pm$  \ph14 \\
1829+29         & 18 29 17.93 & 29 04 58.29 &    9 $\pm$  \ph14 \\
\hline
B3\,VLA         &             &             &                   \\
0809+404        & 08 09 31.62 & 40 28 02.76 &   11 $\pm$  \ph13 \\
0810+460B       & 08 10 58.61 & 46 05 48.12 &    2 $\pm$  \ph12 \\
1025+390B       & 10 25 49.34 & 38 59 57.55 &   46 $\pm$  \ph18 \\
1128+455        & 11 28 56.40 & 45 31 24.57 &    4 $\pm$  \ph13 \\
1159+395        & 11 59 16.35 & 39 35 52.89 &    5 $\pm$  \ph13 \\
1225+442        & 12 25 15.70 & 44 17 17.23 &    1 $\pm$  \ph13 \\
1233+418        & 12 33 10.88 & 41 53 38.00 &   11 $\pm$  \ph14 \\
1242+410        & 12 42 26.39 & 41 04 29.97 &   83 $\pm$     13 \\
1350+432        & 13 50 24.04 & 43 14 09.40 &    4 $\pm$  \ph13 \\
\hline
\end{tabular}
\end{table}

\section{Results and discussion} \label{sec:results}
For each source we have compiled flux densities at different frequencies from 
the literature, mostly from K\"uhr et al. (1979) and from the CATS database
(Verkhodanov et al. 1997). Our new measures at 230~GHz have been added to the 
compilation. All flux densities have been brought to the BGPW scale (Baars et 
al. 1977). Examples of source spectra are shown in Fig.~1.
For the fit algorithm described in Sect. 4.2 we have assumed flux densities less
than 3$\sigma$ to be upper limits only.

Most of the sources show significant departure from the classical power law
which describes a zero age transparent synchrotron spectrum
from a relativistic electron population with power law energy distribution.
The deviations from the power law are of the following type: a) a low-frequency
turnover (the most conspicuous deviation); b) a steepening at high frequencies.
High-frequency flattening, if any, is quite rare. 
The above deviations are interpreted as due to synchrotron self-absorption and to 
particle energy losses, respectively.
In order to describe them, we have fitted the compiled flux densities with a 
synchrotron aged spectrum ${\rm S}_{\rm aged}(\nu)$
(described in the next section), modified by low-frequency absorption, as 
follows (Pacholczyk 1970):
\begin{equation}
{\rm 
S}(\nu)\propto(\nu/\nu_1)^{\alpha+\beta}(1-e^{-(\nu/\nu_1)^{-(\alpha+\beta)}}) 
\cdot {\rm S}_{\rm aged}(\nu) 
\label{a}
\end{equation}
where $\nu_1$ is the frequency at which the optical depth is equal to 1. In 
case of an homogeneous synchrotron self-absorbed source $\beta = 2.5$, while
$\alpha$ is the not aged spectral index in the transparent frequency range. 

\subsection{The synchrotron aged spectrum model}
We assume that the radio source evolution is described by a 
{\it continuous injection model}, where the sources are continuously 
replenished by a constant flow of fresh relativistic 
particles with a power law energy distribution, with exponent $\delta$. 
It is well known that, 
under these assumptions, the radio spectrum has a standard shape 
(Kardashev 1962), with spectral index 
$\alpha_{\rm inj} = (\delta -1)/2$ below a critical frequency $\nu_{\rm br}$ and 
$\alpha_{\rm h} = \alpha_{\rm inj} + 0.5$ above $\nu_{\rm br}$. If there is no 
expansion and the magnetic field is constant, the frequency 
$\nu_{\rm br}$ (in GHz) depends on the elapsed time since the source formation, 
$\tau_{\rm syn}$ (in Myrs), the intensity of the magnetic field $B$ (in $\mu$G) 
and the magnetic field equivalent to the microwave background $B_{\rm CMB} = 
3.25 (1+z)^2$ (in $\mu$G) as:
\begin{equation}
\tau_{\rm syn} = 1610 {\frac{B^{0.5}}{B^2 + B^2_{\rm CMB}}}
{\frac{1}{[\nu_{\rm br}(1+z)]^{1/2}}}
\label{b}
\end{equation}
The whole spectral shape cannot be described by an analytic equation,
the two behaviours described being only the asymptotical ones, and has
to be computed numerically. This model is referred as continuous
injection (CI). Fitting the spectral data to the numerically computed CI spectrum, one 
obtains the break frequency $\nu_{\rm br}$, from which
 the source age is obtained if the magnetic field is known. \par
\noindent
This simple model does not consider expansion effects, which may be
important if the source is young. 
So, the simplest variant of the original model is the one in which the radiating
particles loose energy through expansion and the magnetic field changes 
according to flux conservation (Kardashev 1962, CIE). 
An alternative possibility to be considered, in an expanding source, is that the 
magnetic field changes less rapidly than in the flux conserving assumption 
because of a continuous magnetic flux input associated to the fresh particles 
injection. We set $B \propto t^{-{\rm m}}$, where for m = 2 we get the flux 
conserving expansion. This is referred as CIm model.  
In our special case we assume m = 1, consistent with the models
applied by  Baldwin (1982) or Begelman (1996). 
Although the theoretical background to these models is in the Kardashev
paper, we have decided to present in the Appendix the detailed development.
The spectral shapes of the CIE and CIm models have been computed again 
numerically. The break frequency for the CIE model is 
sixteen times higher than in the CI case, for equal elapsed time and
final magnetic field intensity (Kardashev 1962). In the CIm model, 
instead, one finds that the break frequency is $(2.5{\rm m}-1)^2$ larger than in 
the CI model. The asymptotic behaviours at
frequencies lower and larger than the break frequency are the same
as in the CI model.
However the steepening occurs over a broader frequency interval. 
In order to
better emphasize the differences between these three models it is useful
to consider, together with the usual flux-frequency representation,
 the point-to-point spectral index defined as
\begin{equation}
\alpha(\nu/\nu_{\rm br})=-\frac{d\log {\rm S}_{\rm aged}(\nu/\nu_{\rm
br})}{d\log \nu/\nu_{\rm br}}
\end{equation}
Both representations are shown in Fig.~2. While in the flux-frequency
plane the differences are hardly visible, they can be much better
traced in the point-to-point spectral index behaviour. 
Note that the displayed figures only show the theoretical differences. 
In practice the observed spectra permit only fits in the flux-frequency plane.

\begin{figure}[t]
\epsfxsize=8cm
\epsfbox  [45 150 540 720] {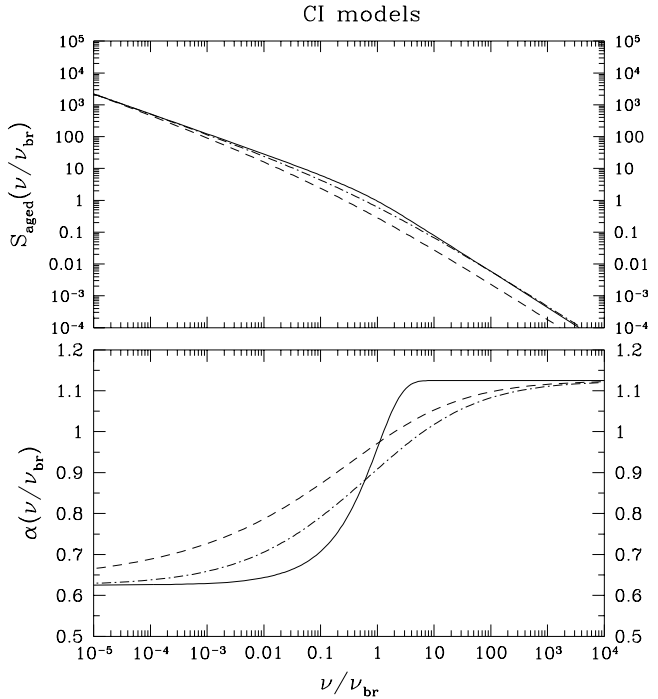}
\caption[]{The three continuous injection models described in the text.
The solid line corresponds to the CI, the dashed line to the CIE and the
dot-dashed line to the CIm. On the top the flux density (arbitrary units)
is plotted as a function of the ratio $\nu/\nu_{\rm br}$ (flux-frequency plane),
on the bottom the point-to-point spectral index is shown as a function of
$\nu/\nu_{\rm br}$. The injection spectral index is the
same for all three models $\alpha_{\rm inj}=0.625$.}
\end{figure}
\subsection{The spectral fits}
Spectral fits to the spectra have been made with the CI, CIE, and CIm
models. They allow us to determine the non-aged spectral index $\alpha_{\rm 
inj}$ 
and the break frequency $\nu_{\rm br}$, that, together with the
normalization, are the three free parameters characterizing all the
models. The best fits are, surprisingly,
obtained with the CI model. The other models have steepenings which are too 
gradual for the majority of the spectra. An example is given in Fig.~3.
The reduced $\chi^2_{\rm red}$ of the
models including adiabatic expansion and a variable magnetic field 
is always greater (typically twice) than that of the CI model (Fig.~4).
The CI fits appear quite good, even in cases of high $\chi^2_{\rm red}$ values,
which, at visual inspection, appear more due to an under-estimation of the flux
density errors than to a poor fit of the spectral model on the data. 
\begin{figure}[t]
\epsfxsize=8cm\epsfbox  [45 150 540 720] {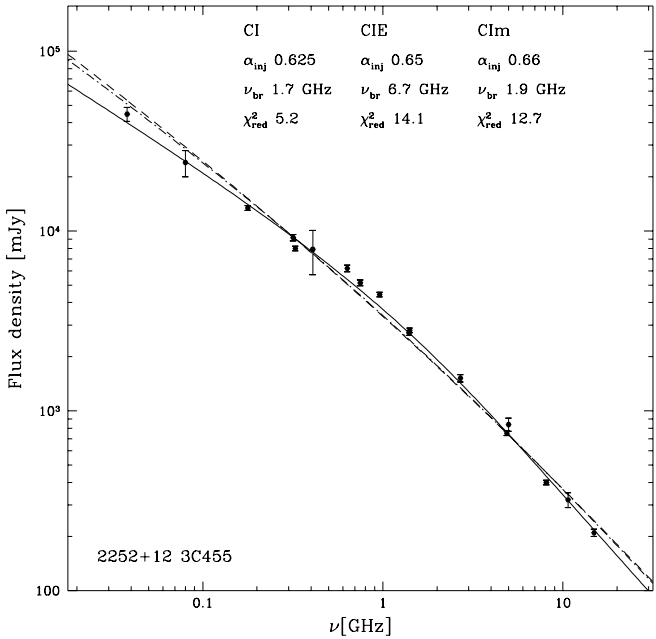}
\caption[]{Three different fit results for the source 2252+12. The solid line
corresponds to a pure CI model, the dashed line represents the CIE model, the
dot-dashed line stands for the CIm model.
Note the different qualities of the fits, expressed by the various
$\chi^2_{\rm red}$ values!}
\end{figure}

The majority of the spectra show a clear break frequency, with a change of slope
$\Delta \alpha \sim 0.5$. We stress that there is no evidence for spectral
steepening with $\Delta \alpha$ significantly larger than 0.5. Only a few
sources are fitted by simple power laws. In these cases $\nu_{\rm br}$ could be
either very high ($\ga 100$ GHz) or very low ($\la {\rm few}~100\;{\rm MHz}$).
In these sources we have preferred the low frequency choice, since for the
high frequency one would have implied abnormally high values for
$\alpha_{\rm inj}$ as compared to the other sources.

We also note that only a very small fraction of the CSS sources, if any,
shows some evidence of flux density excess at high frequency, as it would
be caused by a flat spectrum core or by thermal dust emission. Perhaps
the only case is 3C138, where the core, known from VLBI observations,
shows up in the integrated spectrum at 230 GHz only.

It may appear surprising that the fits with the assumed spectral
model are so good. In fact the sources of the sample consist of several 
components, as lobes and jets and hot spots, where physical conditions 
can differ from one another and therefore also break frequencies may be 
different. It is likely  that the spectrum is dominated by the brighter 
component(s).
In addition, one could imagine that some confusion might have occurred between 
genuine spectral steepening due to energy losses and the low-frequency turnover
due to absorption processes. We feel that this is a minor problem, but it 
is difficult to quantify it (see, however, next section). 

The break frequencies range from a few hundred MHz to tens of GHz. At low 
frequencies, as said above, the limit is set by confusion with the effects of  
absorption processes. The injection spectral index $\alpha_{\rm inj}$
ranges from 0.35 to 0.8, with $\langle \alpha_{\rm inj} \rangle = 0.63$. The typical errors of 
the break frequencies and the injection spectral indices as given by the
fit algorithm are up to 40\% and 0.05, respectively. The results of 
the CI fits are compiled in Tab.~2. 
\begin{table*}
\caption{ Physical Parameters }
\centering
\begin{tabular}{|llrrcrrrc|}\hline
\multicolumn{1}{|c}{Source}&\multicolumn{1}{c}{redshift}& \multicolumn{1}{c}{LS} 
& \multicolumn{1}{c}{$\nu_{\rm br}$}& \multicolumn{1}{c}{$\alpha_{\rm inj}$} & 
\multicolumn{1}{c}{$B_{\rm eq}$} & \multicolumn{1}{c}{$\tau_{\rm syn}$} 
&\multicolumn{1}{c}{$ v_{\rm exp}$/c} & \multicolumn{1}{c|}{type}\\
                &        & \multicolumn{1}{c}{[kpc]}& \multicolumn{1}{c}{[GHz]}  
      &      & \multicolumn{1}{c}{[$10^{-3}$G]}& \multicolumn{1}{c}{[$10^{3}$ yrs]}   &  
& \\
\hline
 3C-PW          &       &       &        &      &       &          &       &    
\\
 0127+23 (3C43)  &  1.46 &  9.40 & $>$100\phu & 0.76 &  5.0  &   0.3
&  55.0\ph1  &  b \\
 0134+32 (3C48)  &  0.37 &  2.10 &  1.8   & 0.50 &  2.0  &  11.3    &  0.60 & b 
\\
 0138+13 (3C49)  &  0.62 &  3.60 &  2.6   & 0.57 &  7.0  &   1.3    &  4.39 & b  
\\
 0221+67 (3C67)  &  0.31 &  6.80 &  2.2   & 0.56 &  0.7  &  51.0    &  0.22 &  a 
\\ 
 0223+34         &  1.00 &  3.80 & 24.1   & 0.38 &  9.4  &   0.3    & 24.20 & b 
\\
 0316+16         &  1.00 &  1.20 & 12.0   & 0.81 &  9.8  &   0.3    &  5.74 & b 
\\
 0345+33 (3C93.1)&  0.24 &  1.20 &  1.6   & 0.58 &  1.0  &  35.8    &  0.05 &  ? 
\\
 0404+76         &  0.60 &  0.53 & 10.7   & 0.38 &  4.9  &   1.1    &  0.76 & a 
\\
 0428+20         &  0.22 &  0.45 &  7.2   & 0.38 &  5.8  &   1.2    &  0.59 & b 
\\
 0429+41 (3C119) &  1.02 &  0.75 &  9.5   & 0.49 &  8.0  &   0.5    &  2.36 & b 
\\
 0518+16 (3C138) &  0.76 &  2.90 &  5.3   & 0.47 &  1.0  &  16.7    &  0.28 &  b 
\\
 0538+49 (3C147) &  0.55 &  2.40 &  1.6   & 0.44 &  2.5  &   8.3    &  0.47 & b 
\\
 0740+38 (3C186) &  1.06 &  8.20 &  0.3   & 0.75 &  0.7  & 113.2    &  0.12 &  a 
\\
 0758+14 (3C190) &  1.20 & 14.10 & 30.3   & 0.79 &  0.7  &  10.6    &  2.15 & b 
\\
 1005+07 (3C237) &  0.88 &  4.50 &  2.4   & 0.59 &  1.2  &  18.1    &  0.40 &  a 
\\
 1019+22 (3C241) &  1.62 &  2.80 &  1.1   & 0.77 &  2.2  &   9.3    &  0.49 &  a 
\\
 1153+31         &  1.56 &  3.20 &  6.9   & 0.69 &  1.7  &   5.4    &  0.95 & a 
\\
 1203+64 (3C268.3)& 0.37 &  3.90 &  4.4   & 0.66 &  0.7  &  35.4    &  0.18 & a 
\\
 1225+36          & 1.98 &  0.17 &  3.9   & 0.64 & 18.0  &   0.2    &  1.41 & b 
\\
 1250+56 (3C277.1)& 0.32 &  4.40 &  0.7   & 0.39 &  0.4  & 204.3    &  0.04 & a 
\\
 1323+32         & 0.37 &  0.18 &  6.9   & 0.38 &  5.1  &   1.4    &  0.20 & a 
\\
 1328+30 (3C286) & 0.85 & 14.20 &  7.3   & 0.38 & 13.0  &   0.3    & 77.82 & b 
\\
 1328+25 (3C287) & 1.06 &  0.40 & 10.7   & 0.47 &  3.5  &   1.7    &  0.39 & b 
\\
 1358+62         & 0.43 &  0.16 & 71.9   & 0.70 &  9.0  &   0.2    &  1.39 & b 
\\
 1416+06 (3C298) & 1.44 &  9.10 & $<$0.1 & 0.50 &  1.6  & $>$50\phu    
&$<$0.3\ph1 & a  \\
 1443+77 (3C303.1)& 0.27 &  5.00 &  0.8   & 0.64 &  0.6  & 110.2    &  0.07 & a 
\\  
 1447+77 (3C305.1)& 1.13 &  9.00 &  0.8   & 0.62 &  0.4  & 151.3    &  0.10 & a 
\\  
 1458+71 (3C309.1)& 0.90 &  7.80 &109.4   & 0.65 &  7.0  &   0.6    & 21.0\ph1  
&  b \\  
 1517+20 (3C318)  & 0.75 &  7.80 &  4.8   & 0.69 &  1.5  &   9.6    &  1.32 & b 
\\  
 1607+26          & 0.47 &  0.18 &  6.3   & 0.71 & 10.0  &   0.5    &  0.55 & a 
\\  
 1634+62 (3C343)  & 0.99 &  0.70 &  3.8   & 0.65 &  2.0  &   6.5    &  0.17 & ? 
\\  
 1637+62 (3C343.1)& 0.75 &  1.30 &  1.7   & 0.62 &  1.5  &  16.0    &  0.13 & a 
\\  
 1819+39          & 0.40 &  3.10 &  3.3   & 0.73 &  6.0  &   1.6    &  3.13 & b 
\\  
 1829+29          & 0.60 &  9.30 & 12.1   & 0.63 &  6.0  &   0.8    & 19.12 & b 
\\  
 2248+71 (3C454.1)& 1.84 &  6.60 &  0.4   & 0.69 &  1.5  & $\ga$26.5& $<$0.40 &  
a \\  
 2249+18 (3C454)  & 1.76 &  2.10 & 11.9   & 0.72 &  5.0  &   0.8    &  4.29 &  b 
\\  
 2252+12 (3C455)  & 0.54 & 13.70 &  1.7   & 0.62 &  0.2  & 348.2    &  0.06 & a 
\\ 
 2342+82          & 0.74 &  0.66 & 10-100 & 0.79 &  4.5  & $\la$1.3 &  0.82 & a 
\\ 
\hline
 B3\,VLA         &        &       &         &         &      &       &       &   
\\
 0809+404        &  0.55 &  4.80 &    4.8  &    0.53 &  1.1 & 16.1  &  0.48 & ? 
\\  
 0810+460B       &  0.33 &  2.80 &    13.9 &    0.94 &  1.1 & 10.3  &  0.44 & a 
\\
 1025+390B       &  0.361&  5.10 &$>$100\phu&   0.65 &  0.8 &  6.0  &  1.40 & a? 
\\
 1128+455        &  0.40 &  1.90 &    1.8  &    0.53 &  1.5 & 17.3  &  0.18 & ?  
\\
 1159+395        &  2.37 &  0.35 &    2.8  &    0.38 &  4.7 &  1.6  &  0.35 &  ? 
\\
 1225+442        &  0.22 &  0.50 &    2.5  &    0.67 &  0.8 & 40.3  &  0.02 &  ? 
\\  
 1233+418        &  0.25 &  2.70 &    3.5  &    0.51 &  0.6 & 52.7  &  0.08 & a? 
\\
 1242+410        &  0.811&  0.40 &    4.7  &    0.38 &  2.6 &  4.2 &  0.16 &  ? 
\\  
 1350+432        &  2.149&  8.00 &    0.2  &    0.84 &  1.2 & 55.4  &  0.23 &  ? 
\\  
\hline
\end{tabular}
\end{table*}
\begin{figure}[t]
\epsfxsize=8cm\epsfbox  [45 150 540 720] {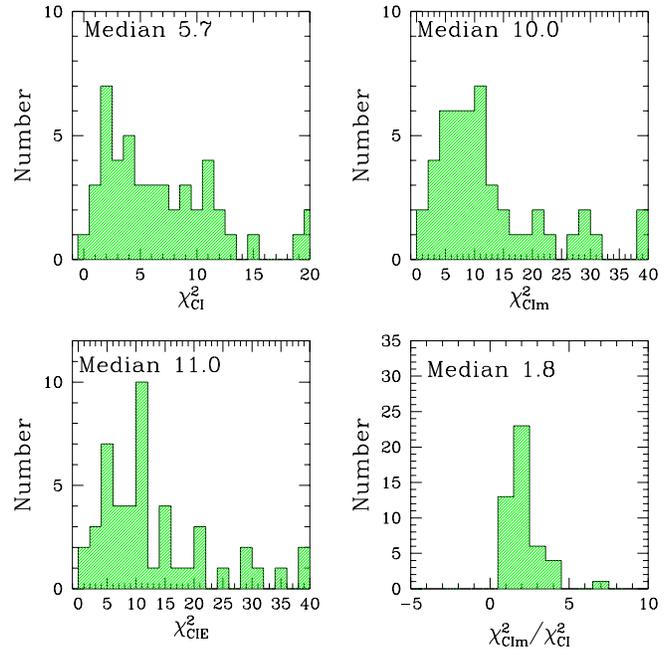}
\caption[]{$\chi^{2}_{\rm red}$ histograms for the CI, CIE, and CIm models. On 
the
bottom right the histogram of the ratio between the CIm and the CI
$\chi^{2}_{\rm red}$ is also shown.}
\end{figure}

\subsection{Radiative ages and the nature of CSS sources}
In order to determine radiative ages, from Eq. (\ref{b}) and variants of the other 
models, the magnetic field $B$ has to be known. 
We stress that the age depends rather strongly on $B$, which is
somewhat uncertain. We take the equipartition value $B_{\rm eq}$. Our assumption
is motivated by the fact that $B_{\rm eq}$ accounts rather well for the
low-frequency turnover in terms of synchrotron self-absorption.
 We are aware that this is not a proof for equipartition.
 Other authors (e.g. Bicknell et al. 1997) prefer instead 
thermal absorption. In any case, $B_{\rm eq}$ represents a poor statistical
upper limit to $B$, in the sense that, were it larger by a factor of four, 
the low frequency turnovers would be systematically higher than observed 
by $\approx$ 30\%.

Using the value of $B_{\rm eq}$, we obtain, from Eq. (\ref{b}), radiative
ages $\tau_{\rm syn}$ ranging from $10^3$ to $10^5$ years. Since the intrinsic
magnetic fields of the CSS sources in our sample strongly overweight the 
magnetic field equivalent to the cosmic microwave background ($B_{\rm CMB}$), 
the latter can be neglected in Eq. (\ref{b}). Therefore, if the source magnetic 
field deviates by a factor {\it f} from the field determined for equipartition 
$B_{\rm eq}$, the radiative ages will change by $f^{-3/2}$.
These radiative ages do not necessarily represent the source ages, but rather
the radiative ages of the dominant source component(s). Only when the 
lobes, which have accumulated the electrons produced over the source lifetime, 
dominate the source spectrum, the radiative age $\tau_{\rm syn}$ is
likely to represent the 
age of the source. 
If, instead, the spectrum is dominated by a jet or by hot spots, 
the radiative age likely represents the permanence time of the electrons in that
component and is expected to be less (perhaps much less) than the source age.
In addition, dominant jets or hot spots might have their break frequency 
up-shifted by Doppler effects. 

The existing structure information on our sample, mostly from MERLIN and VLBI 
observations, allows us to split the sources in two groups: those in which 
the overall spectrum is dominated by lobes (classified ``type a'' in Tab.~2);  those 
in which the spectrum is dominated by a bright jet or hot spot
 (classified ``type b'' in Tab.~2). For the B3\,VLA sample the
available information does not yet allow such a morphological sub-division.

The sources of class b have radiative ages systematically lower than those of 
class a, as we expected.  Furthermore, while the synchrotron
age is well correlated with the source size for class a, it
seems that there is no correlation at all between the linear size and
the radiative age for  class b sources (see Fig.~5).
\begin{figure}[t]
\epsfxsize=8cm\epsfbox  [45 150 540 680] {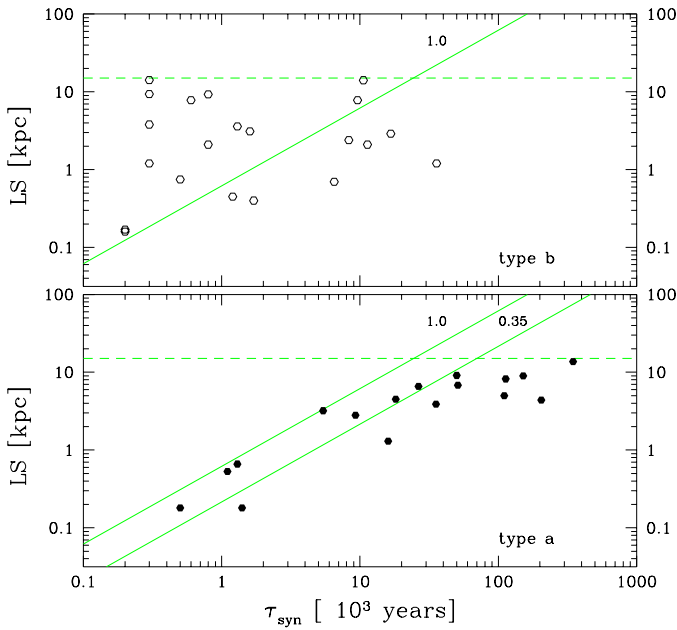}
\caption[]{Linear size as a function of the synchrotron age for type a
(filled dots) and type b (open dots) sources. The B3\,VLA sources have
been excluded. The horizontal dashed lines represent the selection limits
of the linear size distribution of the sources in our samples while the 
diagonal lines reflect constant values of $\frac{v_{\rm exp}}{c}$.}
\end{figure}
We have further computed for each source the expansion velocity $v_{\rm exp} 
\approx {\rm LS}/2 \tau_{\rm syn}$, where LS is the source largest dimension.
The two classes show very different distributions:\\
Class a $\langle v_{\rm exp}/c \rangle = 0.34 \pm 0.07~~~ \sigma_{v_{\rm exp}} = 0.28$\\
Class b $\langle v_{\rm exp}/c \rangle = 11 \pm 4.4~~~ \sigma_{v_{\rm exp}} = 20$\\
Provided the assumed magnetic field is reasonably correct, the above values
for the radiative ages indicate that {\it the CSS sources are young}.
We further note that the
ages, and corresponding expansion velocities, are not far (somewhat larger)
from the recent results on expansion of CSOs by Owsianik et al. (1998)
and Owsianik \& Conway (1998).
As the radiative ages are strongly dependent on the assumed magnetic field,
a field only a factor two lower than assumed would be required for a
better agreement.
\par \noindent
In order to maintain the {\it frustration scenario}, in which the sources'
lifetimes are $\approx 10^7$ years, their equipartition magnetic field
should be decreased by a factor $\ga 20$. 
\par \noindent
One could think that the correlation between linear
sizes  and the radiative ages shown in Fig.~5 could be a partial consequence
of the equipartition assumption. In fact, $B_{\rm eq} \propto {\rm LS}^{-6/7}
$ implies that ${\rm LS} \propto \tau_{\rm syn}^{7/9}$.
This is not the case for the following reasons: 1) no correlation between
linear sizes and radiative ages is obvious for type b sources. \par \noindent
 2) In particular, the break frequencies seem to be correlated with the
linear sizes for type a sources (Fig.~6). \par \noindent
These means, at least for class a sources, that the break frequency is an
effective clock indicating the source age. The correlation seen in Fig.~5 is 
not an artifact completely introduced by the assumption of equipartition in 
Eq. (2). \par \noindent

\begin{figure}[t]
\epsfxsize=8cm
\epsfbox  [45 150 540 680]{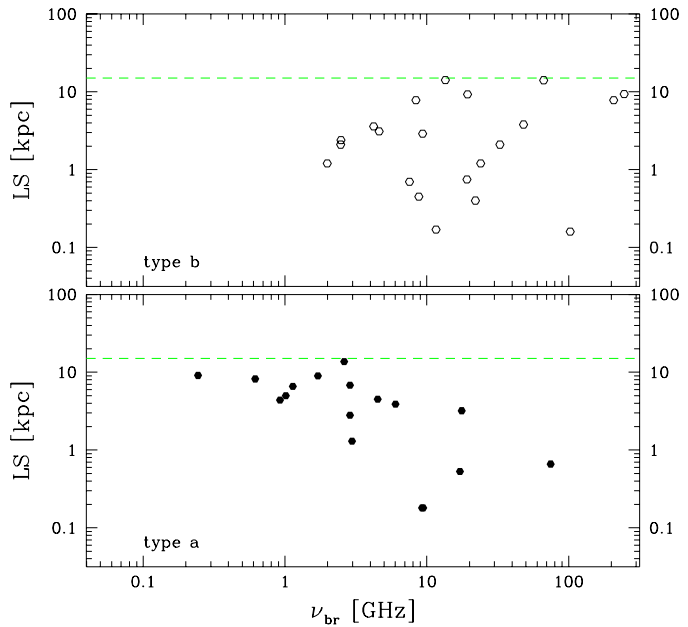}
\caption[]{Linear size as a function of the break frequency (in the source
rest frame) for type a
(filled dots) and type b (open dots) sources. The B3\,VLA sources have
been excluded. The horizontal dashed lines represent the selection limits
of the linear size distribution of the sources in our samples.}
\end{figure}

\section{Conclusions}
The high-frequency integrated spectrum of CSS sources 
shows a break with a moderate spectral steepening ($\Delta \alpha \approx 0.5$),
well fitted by the continuous injection model spectrum with constant
magnetic field. Spectral fits with synchrotron models with decreasing magnetic 
field are definitely poorer.
There is no evidence of sharp cut-offs as would be expected if the supply of
relativistic electrons had stopped more than $\approx 10^4$ years ago.

In lobe dominated CSS sources the radiative ages $\tau_{\rm syn}$ are likely to 
represent 
the source ages. Assuming equipartition magnetic field strengths the source ages are
in the range of up to $10^5$ years, in agreement with the recent results on 
expansion velocities of small size CSOs by Owsianik et al. (1998)
and Owsianik \& Conway (1998) derived from VLBI observations.
Therefore, assuming the CSS sources to be young, the magnetic field has
been deduced to be equal to the equipartition value within a factor $\approx 2$.

In jet or hot spot dominated CSS sources, the radiative life times are much 
shorter and very likely represent the permanence time of the electrons 
in those components.

\begin{acknowledgements}
We are grateful to Drs. B. Cotton and S. Spangler for very
helpful comments on the manuscript of this paper and to Dr. H.
Andernach for the suggestions on the compilation of the flux densities.
We made use of the database CATS (Verkhodanov et al. 1997) of 
the Special Astrophysical Observatory.
We acknowledge the Italian Ministry for University and Scientific
Research (MURST) for partial financial support (grant Cofin98-02-32).
Part of this work was supported by European Commission, TMR Programme, 
Research Network Contract ERBFMRXCT97-0034 ``CERES''.    
\end{acknowledgements}
\newpage
\appendix{\bf Appendix}
\section{Synchrotron radiation, adiabatic expansion and continuous
injection in the case of a time-variable magnetic field}
We assume that adiabatic expansion  and injection of particles start
at time $t_{0}$. The source is
continuously replenished by a constant flow of particles with a power
law energy spectrum with spectral index $\delta$
intensity $B$ decreases according to
$$ B=B_{0}\left(\frac{t_{0}}{t} \right)^{m} $$
and the characteristic size $R$ of the regions occupied by the
particles (which should not be confused with the source size LS) grows as
$$R=R_{0}\left(\frac{t}{t_{0}}\right)^{r}$$	
The continuity equation for this model is 
\begin{equation}
\frac{\partial N}{\partial t}=a_{1} \frac{\partial}{\partial
\epsilon}(\epsilon^{2} N) +a_{2} \frac{\partial}{\partial
\epsilon}(\epsilon N)+a_{3} \epsilon^{-\delta}
\label{ciecontinuity}\end{equation}
$N(\epsilon,t)$ is the differential energy spectrum of the
relativistic particles at time t.
The first term  on the right hand side describes
the effects of the synchrotron losses, $a_{1}=c_{1}B_{\perp}^{2}$, $c_{1}$ is constant (Pacholczyk 1970). The second term 
 accounts for adiabatic expansion ($a_{2}=r/t$).
The term  $a_{3}$ describes the injection rate and is considered to be constant for simplicity.\\
The solution of Eq. (\ref{ciecontinuity}) is
\begin{eqnarray} 
N(\epsilon,t)&&=a_{3}\epsilon^{-\delta} \cdot \nonumber \\
&&\int_{t_{\rm low}}^{t}\left( \frac{\tau^{\prime}}{t} \right)^{r(\delta-1)}\left( 1-\frac{ \epsilon}{{\epsilon^{*}_{\rm br}(\tau^{\prime})}} \right) ^{\delta-2}d\tau^{\prime}
\label{CIRMsolution}
\end{eqnarray}
where $t_{\rm low}$ is MAX$[t_{0},\tau_{\rm min}]$ and $\tau_{\rm min}$ is implicitly given by
\[\epsilon=\frac{2m+r-1}{ c_{1} B_{\perp}^{2} t
\left[(\frac{t}{\tau_{\rm min}})^{2m+r-1}-1
\right]}\]
The energy ${\epsilon^{*}_{\rm br}}(\tau)$ is
\[ {\epsilon^{*}_{\rm br}}(\tau)=\frac{2m+r-1}{ c_{1} B_{\perp}^{2} t( (t/\tau)^{2m+r-1}-1)}
\label{ciebreaktau}\]
$\epsilon^{*}_{\rm br}(\tau)$ is the break energy of the particle population
injected at the time $\tau$ ( $t_{0} \le \tau \le t$).
The solution (\ref{CIRMsolution}) represents the sum of all the
populations injected form $\tau=t_{0}$ to $\tau=t$.
The integral (\ref{CIRMsolution}) can be written as
\begin{eqnarray}
N(\epsilon,t)&&=a_{3}\epsilon^{-\delta-1}t \frac{{\epsilon_{\rm br}}}{2m+r-1} \cdot \nonumber \\
&& \int_{z_{\rm low}}^{1}\left[(1-z)\frac{{\epsilon_{\rm br}}
}{\epsilon}+1\right]^{-\frac{r \delta+2m}{2m+r-1}}   
z^{\delta-2}dz
\label{CIRMzsolution}
\end{eqnarray}
where $z_{\rm low}$ is equal to MAX$[0,1-\epsilon/ {\epsilon_{\rm br}}_{\rm min}]$,
\[ {\epsilon_{\rm br}}=\frac{2m+r-1}{ c_{1}B_{\perp}^{2} t }
\label{CIRMbreak}\]
and
\[ {\epsilon_{\rm br}}_{\rm min} = \frac {
{\epsilon_{\rm br}}} {( t/t_{0}) ^{2m+r-1}-1 }\]
In the energy spectrum two break energies are  always present,  
${{\epsilon_{\rm br}}}$ and ${{\epsilon_{\rm br}}_{\rm min}}$, 
 the break energy of the first population injected at $t=t_{0}$ .
These regions can be identified in the energy spectrum:
\begin{itemize}
\item[1)]$\epsilon \ll {\epsilon_{\rm br}}_{\rm min}$ .\\
 In this region the spectrum is a power law:
 \[ N(\epsilon,t)\simeq \frac{ a_{3}\epsilon^{-\delta} t}{r(\delta-1)+1}\left[
1-\left(\frac{t_{0}}{t}
\right)^{r(\delta-1)+1}\right]\]
\item[2)] $ {\epsilon_{\rm br}}_{\rm min} \approx \epsilon < {\epsilon_{\rm br}}$.\\
The spectrum starts to gradually deviate from the low energy power law 
at  $\epsilon \approx  {\epsilon_{\rm br}}_{\rm min}$ and undergoes a
steepening at
 $\epsilon \approx  {\epsilon_{\rm br}}$.\\
\begin{eqnarray*}
N(\epsilon,t)&=&a_{3}\epsilon^{-\delta-1}t \frac{{\epsilon_{\rm br}}}{2m+r-1} \cdot  \\ 
&&\int_{z_{\rm low}}^{1}\left[(1-z)\frac{{\epsilon_{\rm br}}}{\epsilon}+1\right]^{-\frac{r \delta+2m}{2m+r-1}} z^{\delta-2}dz 
\end{eqnarray*}
\item[3)]$ \epsilon \gg {\epsilon_{\rm br}}$.\\
The spectrum is again a power law but with a spectral index steeper by 1.
\[N(\epsilon,t) \simeq \frac{ a_{3}\epsilon^{-\delta-1}}{c_{1} B_{\perp}^{2}(\delta-1)}\]
\end{itemize}
Since the separation between these two energies increases with time
the shape of the spectrum also changes with time. But an asymptotic
stationary shape still exists when $t/t_{0} \rightarrow \infty $. In
fact, in this case $ {\epsilon_{\rm br}}_{\rm min} \rightarrow 0 $,  and at every
energy $\epsilon$, $z_{\rm low}=0$. In this way the integral
\begin{eqnarray*}
N(\epsilon,t)&=&a_{3}\epsilon^{-\delta-1}t \frac{{\epsilon_{\rm br}}}{2m+r-1} \cdot \\  
&&\int_{0}^{1}\left[(1-z)\frac{{\epsilon_{\rm br}}}{\epsilon}+1\right]^{-\frac{r \delta+2m}{2m+r-1}} z^{\delta-2}dz
\end{eqnarray*}
does not depend on the ratio $t/t_{0}$ anymore.\\
The integral can be solved only by numerical means, but
analytic asymptotic limits can still be found.
\begin{eqnarray}
N(\epsilon,t) \simeq \frac {a_{3} \epsilon^{-\delta} t}{r( \delta-1)+1 } 
~~~~~~~~~~ \epsilon \ll \frac{2m+r-1}{c_{1} B_{\perp}^{2} t} \nonumber \\
N(\epsilon,t) \simeq \frac {a_{3}\epsilon^{-\delta-1}} {
c_{1}B_{\perp}^{2}(\delta-1)} ~~~~~~~~~~\epsilon \gg \frac{2m+r-1}{c_{1} B_{\perp}^{2} t}
\label{CIRMlimits}
\end{eqnarray}
i.e. in the spectrum  only the ${\epsilon_{\rm br}}$ break
energy is present.
The asymptotic solutions are two power laws. The low-energy power law
spectral index is $\delta$, the high-energy power law spectral index is
$\delta+1$. Moreover, the normalization for  $\epsilon \gg
{\epsilon_{\rm br}}$ is time-independent, i.e. 
a perfect balance is reached between the number of particles that
leave this region because the synchrotron  and expansion losses and the number of
particles constantly injected into the source.\\ 
\subsection{CI model}
The CI model reproduces the simple situation in which there is no
adiabatic expansion and the magnetic field strength stays constant. 
This corresponds to $r=0$ and $m=0$ in (\ref{CIRMzsolution}).
In this case the integral can be directly solved resulting in:
\begin{eqnarray*}
N(\epsilon,t)&=&\frac{a_{3}\epsilon^{-\delta-1}}
{c_{1}B_{\perp}^{2}(\delta-1)} \left[ 1- \left(1-\epsilon/{\epsilon_{\rm br}}
\right) ^{\delta-1} \right]~~~~~~\epsilon<{\epsilon_{\rm br}} \\
N(\epsilon,t)&=&\frac{a_{3}\epsilon^{-\delta-1}}
{c_{1}B_{\perp}^{2}(\delta-1)} ~~~~~~~~~~~~~~~~~~~~~~~~~~~~~~~~~~~~\epsilon \geq {\epsilon_{\rm br}} \nonumber
\label{cisolution}
\end{eqnarray*}
where
\[{\epsilon_{\rm br}}=\frac{1}{c_{1}B_{\perp}^{2}(t-t_{0})}\]
The shape of the spectrum does not depend on time (Kardashev 1962).
\subsection{CIE model}
The CIE model reproduces the situation in which the volume containing
the particles is adiabatically expanding at a
constant rate and the magnetic field is frozen in the plasma. Since $R \propto t$, the conservation of the magnetic flux
requires that $B \propto t^{-2}$.
 This corresponds to $r=1$ and $m=2$ in (\ref{CIRMzsolution}).
The shape of the spectrum changes with time during its rise. However an
asymptotic stationary solution exists when $t \gg\ t_{0}$:
\[N(\epsilon,t) \simeq \frac{a_{3}\epsilon^{-\delta-1}t}{c_{1}B_{\perp}^{2}}\int_{0}^{1}\left[(1-z)\frac{{\epsilon_{\rm br}}}{\epsilon}+1\right]^{-\frac{\delta}{4}-1} z^{\delta-2}dz\]
\[{\epsilon_{\rm br}}=\frac{4}{c_{1}B_{\perp}^{2}t}\]
In this asymptotic limit the CIE spectrum shows a break energy four
 times greater than the break energy of the CI model. Using
(\ref{CIRMlimits}) one finds that 
\begin{eqnarray*}
N(\epsilon,t) &\simeq& \frac {a_{3} \epsilon^{-\delta} t}{ \delta } 
~~~~~~~~~~~~~~~~~~~~\epsilon \ll \frac{4}{c_{1} B_{\perp}^{2} t} \nonumber \\
N(\epsilon,t) &\simeq& \frac {a_{3}\epsilon^{-\delta-1}} {
c_{1}B_{\perp}^{2}(\delta-1)} ~~~~~~~~~~~~\epsilon \gg \frac{4}{c_{1} B_{\perp}^{2} t}
\label{cielimits}
\end{eqnarray*}
this solution is the same found by Kardashev (1962).
The asymptotic solutions are two power laws with the same dependence on
the spectral index as in the case of the CI model. However, the 
 energy range necessary to complete the transition between the two
asymptotic power laws is wider.
For $t \gg t_{0}$ the break energy of the CIE model is greater than
  the break energy ${\epsilon_{\rm br}}_{\rm min}$ of a CI model with the same $\it{final}$ magnetic field.
This result appears a bit surprising at  first glance. In fact,
due to the expansion losses and the stronger magnetic field, the
break energy of the first population of particle injected at $t_{0}$
in the CIE model is $\it{lower}$ than that of the corresponding
population of the CI model. However, both the stronger mean  magnetic
field  and the expansion losses will also
 decrease the energy of the first populations in a way
that, at time $t$, they do not contribute appreciably to the integrated 
spectra.\\
\subsection{CIm model}
The CIm model reproduces the situation in which  the volume
containing the particles is adiabatically expanding
 and there is always equipartition between magnetic field and particle
energy density. The equipartition condition $u \propto B^{2}$ ($u$ is the energy
density of the relativistic particles) implies that $R \propto
t^{m/2}$. This  corresponds to $r=m/2$ in (\ref{CIRMzsolution}).
An asymptotic stationary solution exists when $t \gg\ t_{0}$:
\[N(\epsilon,t) \simeq
\frac{a_{3}\epsilon^{-\delta-1}t}{c_{1}B_{\perp}^{2}} \int_{0}^{1}\left[(1-z)\frac{{\epsilon_{\rm br}}}{\epsilon}+1\right]^{-\frac{ m \delta+4m}{5m-2}} z^{\delta-2}dz\]
\[{\epsilon_{\rm br}}=\frac{2.5m-1}{c_{1}B_{\perp}^{2}t} \]
In this asymptotic limit the CIm spectrum shows a break energy ($2.5m-1$)
 times greater than the break energy of the CI model. Using
(\ref{CIRMlimits}) one finds that 
\begin{eqnarray*}
N(\epsilon,t) &\simeq& \frac {a_{3} \epsilon^{-\delta} t}{\frac{m}{2}
(\delta-1)+1 } 
 ~~~~~~~~~~~ \epsilon \ll \frac{2.5m-1}{c_{1} B_{\perp}^{2} t} \nonumber \\
N(\epsilon,t) &\simeq& \frac {a_{3}\epsilon^{-\delta-1}} {
c_{1}B_{\perp}^{2}(\delta-1)}  ~~~~~~~~~~~~  \epsilon \gg
\frac{2.5m-1}{c_{1} B_{\perp}^{2} t}
\label{cimlimits}
\end{eqnarray*}
For the CIm the CIE comments are also valid.
\subsection{Emission spectrum}
The emission spectrum is given by the convolution of the emission spectrum of
the single  electron $F(\nu/(c_{2}Bsin\theta \epsilon^{2}))$ (with $c_{2}$ constant, Pacholczyk 1970) with the energy distribution $N(\epsilon,t)$:
\begin{eqnarray*}
 {\rm S}_{\rm aged}(\nu/\nu_{\rm br},\alpha_{\rm inj})& \propto&  \int_{0}^{\frac{\pi}{2}}
\int_{0}^{\infty} F\left(\frac{\nu/\nu_{\rm br}}{x^{2}sin \theta} \right) \cdot \\ 
&& N(x\cdot{\epsilon_{\rm br}} ,t)~ sin ^{2}\theta ~dx ~ d\theta
\end{eqnarray*}
where $\nu_{\rm br} =c_{2} B {\epsilon_{\rm br}}^{2}$ and $\alpha_{\rm inj}=(\delta-1)/2$.
It is assumed that the pitch angles $\theta$ between the electron velocity and the magnetic field direction are isotropically distributed and that the time scale for their continuous re-isotropization is much shorter than the radiative
time-scale, $c_{1} \langle B^{2}_{\perp} \rangle =\frac{2}{3} c_{1}B^{2}$ (Jaffe \& Perola 1974).

\end{document}